# Stochastic dynamic simulations of fast remagnetization processes: recent advances and applications

Dmitri V. Berkov*, Natalia L. Gorn

*INNOVENT e.V., Pruessingstr. 27B, D-07745, Jena, Germany*

**Abstract**

Numerical simulations of fast remagnetization processes using the stochastic dynamics are widely used to study various magnetic systems. In this paper we first address several crucial methodological problems of such simulations: (i) the influence of the finite-element discretization on the simulated dynamics, (ii) choice between Ito and Stratonovich stochastic calculi by the solution of micromagnetic stochastic equations of motion and (iii) non-trivial correlation properties of the random (thermal) field. Next we discuss several examples to demonstrate the great potential of the Langevin dynamics for studying fast remagnetization processes in technically relevant applications: we present numerical analysis of equilibrium magnon spectra in patterned structures, study thermal noise effects on the magnetization dynamics of nanoelements in pulsed fields and show some results for a remagnetization dynamics induced by a spin-polarized current.

*Keywords:* Micromagnetic simulations; stochastic dynamics; colored noise; spin injection
PACS: 75.40.Mg; 75.40.Gb; 75.30.Ds; 72.25.-b;

## 1. Introduction and Motivation

During the last two decades both micromagnetic simulations and experimental methods for studies of fast remagnetization processes made a substantial progress. State of the art of the simulations allows in many cases to achieve a quantitative agreement between simulated and experimental data and to make non-trivial theoretical predictions. In particular the so called Langevin (stochastic) dynamics formalism which allows to include thermal fluctuations into the equation of motion for the magnetization is a powerful tool to study the behaviour of nanosized systems where finite temperature effects play an important role due to small system sizes.

In this review we restrict ourselves to the analysis of simulations performed using the standard Landau-Lifshitz-Gilbert equation of the magnetisation motion

$$\frac{d\mathbf{M}_i}{dt} = -\gamma[\mathbf{M}_i \times (\mathbf{H}_i^{det} + \mathbf{H}_i^{fl})] \\ -\frac{\gamma\lambda}{M_S}[\mathbf{M}_i \times [\mathbf{M}_i \times (\mathbf{H}_i^{det} + \mathbf{H}_i^{fl})]] \quad (1)$$

where the deterministic effective field $\mathbf{H}^{det}$ is augmented by the thermal (fluctuation) field $\mathbf{H}^{fl}$ which is supposed to take into account random fluctuations.

## 2. Methodological problems of dynamical simulations

*2.1. Discretization effects on the remagnetization dynamics for $T = 0$*

Most methodical problems in dynamical micromagnetic simulations arise due to a finite-element representation of a continuous magnetic system.

The first such problem which we briefly recall here is present already for simulations without thermal noise (i.e., $T = 0$ and $\mathbf{H}^{fl} = 0$ in (1)) and is due to the following basic difference between quasistatic and dynamic calculations. For the static case it is sufficient to discretize the system using the mesh size somewhat smaller than the characteristic (exchange or demagnetizing) magnetic length of the material. For dynamics it is, generally speaking, not true. Even if the chosen mesh is fine enough to represent all the magnetization features for the starting state, it might happen that during the remagnetization process magnons with the wavelengths shorter than the grid cell size play an important role. In this case magnons with shorter and shorter wavelengths appear when the remagnetization goes on. As soon as the grid is unable to support these magnons, simulations became inadequate (see detailed analysis in [1]). The effect is particularly important for systems with low dissipation ($\lambda \leq 0.01$) where it poses the *upper* limit on the simulation time accessible for the given grid size.

---

* Corresponding author. Tel.: +49-(0)3641-282537; fax: +49-(0)3641-282530
E-mail address: db@innovent-jena.de

*2.2. Dynamic simulations for T > 0*: *the proper choice of the stochastic calculus*

To simulate the system remagnetization taking into account thermal fluctuations we have to include a random field $\mathbf{H}^{fl}$ into the equation of motion (1). It was shown for a system of non-interacting single-domain particles surrounded by a thermal bath, that in order to reproduce correctly the equilibrium thermodynamics, it is sufficient to take $\mathbf{H}^{fl}$ with all projections having zero mean values and $\delta$-functional correlation properties

$$\langle H^{fl}_{\xi,i} \rangle = 0, \quad \langle H^{fl}_{\xi,i}(0) \cdot H^{fl}_{\psi,j}(t) \rangle = 2D \cdot \delta(t) \delta_{ij} \delta_{\xi\psi} \quad (2)$$

$$D = \frac{\lambda}{1+\lambda^2} \cdot \frac{kT}{\gamma\mu} \quad (3)$$

($i$, $j$ are the particle indices; $\xi$, $\psi = x, y, z$), whereby the noise power $D$ can be evaluated from the fluctuation-dissipation theorem (FDT) [2].

As it was pointed out already by Brown [2], the inclusion of $\mathbf{H}^{fl}$ into Eq. (1) converts it into a stochastic differential equation (SDE): the time evolution of the magnetization includes now a random component making the function $M(t)$ non-differentiable (like a Wiener process) due to correlation properties (2). And the noise in (1) is multiplicative: the random field components $H^{fl}_\xi$ are *multiplied* by the functions to be evaluated ($M_\xi$).

The main difficulty in solving such equations using numerical methods is due to the well known fact that results may depend on the position of the integration points inside the discrete time slices [3]. Usually the so called *Stratonovich* integration (intermediate points in the middle of each time slice) is an adequate choice. However, many widely used numerical methods use the intermediate points at the beginning or the end of a time slice (e.g., the Euler method or implicit methods often favoured for their stability). Such methods converge to the so called *Ito* solution, which, generally speaking, does not reproduce correctly even the *equilibrium* behaviour of a physical system [3].

This problem was extensively discussed in the literature and it was even claimed [4] that micromagnetic simulation results obtained with methods not converging to the *Stratonovich* integral should be discarded. Fortunately, we could show [5, 6] that for standard micromagnetic models with the *constant* moment magnitude of each discretization cell the choice of a stochastic calculus does not matter - *Ito* and *Stratonovich* solutions are equivalent. However, we point out that the proof *heavily relies* on the conservation of the moment magnitude [5]. Hence for models where this is *not* the case - e.g., by simulations of the heat assisted magnetic recording or for models attempting to relax the local restriction $M_i$ = Const [7] - one should pay close attention to the choice of a numerical method used to solve the Eq. (1).

*2.3. Magnetization dynamics simulations for T > 0*: *non-trivial correlation properties of the random field*

The simplest correlation properties of $\mathbf{H}^{fl}$-components given by (2) were derived by Brown [2] for a single magnetic moment in a thermodynamic equilibrium. Hence the question arises whether these properties survive for a finite-element micromagnetic model with complicate interactions between the elements. This question includes two separate problems: (i) whether the random noise in an interacting system remains $\delta$-correlated in space and time and (ii) how to evaluate the noise power $D$ for such a system.

To clarify the first question we start with the statement that the presence of interactions between the moments (cells) by itself does *not* automatically imply non-trivial correlations between the random field components on different cells, because these interactions are already included into the *deterministic* field. Nevertheless there exist at least two reasons why such non-trivial correlations can exist.

The first - physical - reason is the correlation of heat-bath fluctuations responsible for the existence of the random field $\mathbf{H}^{fl}$. These correlations are expected to be very short-ranged both in time and space [8] (correlation time about several ps and correlation length about several nm), so that for standard micromagnetic simulations the assumption about their $\delta$-functional behaviour is at least a good approximation. However, when the remagnetization processes on the time and length scales mentioned above will become a subject of interest (which is expected to happen in the nearest future), it will be necessary to include corresponding correlations into the simulation code.

The second source of non-trivial random field correlations are the above mentioned short-wave magnons which can not be supported by the discretization grid. These magnons can still have a mean free path *much larger* than the grid cell size, thus causing substantial correlations especially of the exchange fields on neighbouring cells. Although these excitations can not be included into simulations on the given grid explicitly, it is possible to take them into account as an additional contribution to the fluctuation field $\mathbf{H}^{fl}$ with the corresponding correlation properties.

To calculate the correlation functions (CF) of the field caused by these magnons one should perform simulations on a much finer grid, then cut out the field components with the wave vectors accessible for the initial (coarse) grid and evaluate the CF of the remaining field [9]. It turns out that the this CF oscillates with a decaying amplitude. Its the initial value, decay time and distance strongly depend on the concrete system. The general advise is to take these correlations into account when the initial value of their CF is comparable with the white noise amplitude in (2).

The second question - about the random noise *power* - is more complicated. Its commonly used value (3) is evaluated for a system of *non-interacting* moments using the FDT which is valid for systems in a *thermodynamical equilibrium*. This result can be easily expanded to a case of *interacting* moments using the transformation of system coordinates to its normal modes. In the harmonic approximation (near equilibrium) such modes can be considered as independent and the power of random noise for all modes is the same (equipartition theorem) and given by (3) (FDT). The backward transformation to the initial coordinates, being - as it was the forward transformation - an orthogonal one, will conserve this noise power value, thus leaving the result (3) unchanged for a system of interacting moments also.

The situation for the *non-equilibrium* remagnetization processes is unfortunately much more complicated. The standard FDT can not be applied to the states passed by the system during such processes, because these states are far from equilibrium. The problem of how to describe correctly the influence of thermal fluctuations during such transitions is now the subject of an intensive research.

*2.4. Influence of a discretization on system properties in a thermodynamical equilibrium*

Finite-element discretization of a continuous micromagnetic system has also important consequences for its equilibrium thermodynamic properties. This problem has not been (up to our knowledge) addressed before, so we shall discuss it here in more detail.

To simplify our discussion we consider the following system: a square region of an extended thin magnetic film in an external field perpendicular to the film plane. We neglect anisotropy and magnetodipolar interactions and use periodic boundary conditions.

One of the most important characteristics of any magnetic system is the power spectrum of its excitations in a thermodynamic equilibrium. Such a spectrum can be efficiently computed using the Langevin dynamics formalism in a following way [6]. We begin with the saturated magnetization state (along the external field) which minimizes the system energy for $T = 0$. Then we integrate the SDEs (1) until the thermodynamic equilibrium is reached - i.e., the total energy does not change systematically with time.

Starting from this moment we save the trajectories of every cell magnetization during the time interval necessary to compute the spectral power with the desired accuracy. Finally we perform the *temporal* Fourier transformation (FT) of these trajectories and averaging of the spectra over several thermal noise realizations.

Typical result of such simulations for a film with lateral sizes 400 x 400 nm, thickness 5 nm, exchange constant $A = 10^{-6}$ erg/cm and $M_S = 1000$ G in an external field $H_{ext} = 100$ Oe is presented in Fig. 1. Here the power spectra of the $m_x$-projection oscillations by $T = 10$ K (to compare with the low-temperature analytical solution given below) for two different discretizations are shown. The most striking features of these spectra are (i) the presence of a sharp cusp in the middle and (ii) a shift of this cusp towards higher frequencies when the discretization is refined.

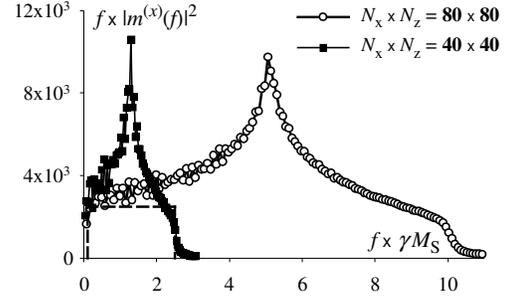

Fig. 1. Power spectrum of $m_x$-oscillations for various discretizations. Cusps are the manifestations of the magnon DoS singularity (Fig. 2). The dashed rectangle represents the spectrum expected in the limit of very fine discretization for a low frequency region available for the 40 x 40 discretization.

This result can be easily understood as follows. The system energy $E = E_{ext} + E_{exch}$ in a continuous formulation is

$$E = -\int_V \mathbf{H}_{ext}\mathbf{M}dV + A\int_V [(\nabla m_x)^2 + (\nabla m_y)^2 + (\nabla m_z)^2]dV \quad (4)$$

After the in-plane discretization of the film into $N_x$ x $N_y$ cells with the sizes $\Delta x$ and $\Delta y$ and volumes $\Delta V$ this energy is converted into the sum over the cell indices $i,j$

$$E = -\mu \sum_{i,j=1}^{N_{x(y)}} \mathbf{m}_{ij}\mathbf{H}_{ij}^{ext} - \frac{1}{2}J\mu^2 \sum_{\langle i,j \rangle}(\mathbf{m}_i\mathbf{m}_j) \quad (5)$$

where $\mu = M_S \Delta V$ is the cell magnetic moment. The exchange constant $J$ in (5) depends on the exchange stiffness $A$ and the grid cell parameters as

$$J = \frac{A}{M_S^2 \cdot \Delta V}\left(\frac{1}{\Delta x^2} + \frac{1}{\Delta y^2}\right) \quad (6)$$

The model described by the energy (5) - which is actually simulated in numerical micromagnetics - is a typical simple *lattice* model which thermodynamic properties are well known. In the small temperatures (large external field) limit one can expand the energy (5) over small in-plane magnetization projections $m_{ij}^x$ and $m_{ij}^y$. Transition to spatial Fourier-components of these projections leads to the energy

$$E = E^{(0)} + \frac{1}{2}N_xN_y \cdot \frac{\mu}{\gamma}\sum_{p,q=1}^{N_{x(y)}} B_{pq} \cdot \omega_{pq} \quad (7)$$

where the eigenfrequencies $\omega_{pq} = \gamma \cdot (H_{ext} + \mu J \cdot f_{pq})$ depend on the wave vector indices $p$ and $q$ via the sum of cos-functions as $f_{pq} = 2 \cdot (2 - \cos(2\pi p/N_x) - \cos(2\pi q/N_z))$. Factors

$$B_{pq} = \langle |m_{pq}^{(x)}|^2 + |m_{pq}^{(z)}|^2 \rangle_T \qquad (8)$$

are given by the thermal average of the squared amplitudes of Fourier harmonics $m_{pq}^x$ and $m_{pq}^y$. In a thermodynamic equilibrium $B_{pq}$ are inversely proportional to eigenfrequencies as $\sim T/\omega_{pq}$, so that the energy contributions of all modes defined by products $B_{pq} \cdot \omega_{pq}$ are equal and proportional to $T$. Hence that the total oscillation power for the given frequency $\omega$ is inversely proportional to this frequency and directly proportional to the number of modes contributing to this frequency, i.e., to the magnon density of states $\rho(\omega)$.

For 2D lattice models with the nearest-neighbours harmonic interaction (as the linearized model (5)) the eigenfrequencies depend on the wave vectors in a cos-like manner (see expression for $f_{pq}$ after (7)) and the density of states (Fig.2) contains the famous van Hove singularity in the middle. It is clearly visible in both spectra in Fig. 1 as a cusp. The spectrum shift towards higher frequencies when the discretization is refined can be easily explained: the eigenfrequencies $\omega_{pq} = \gamma \cdot (H_{ext} + \mu J \cdot f_{pq})$ are proportional to the exchange constant $J$ which according to (6) increases as an inverse square of a cell size when a mesh is refined.

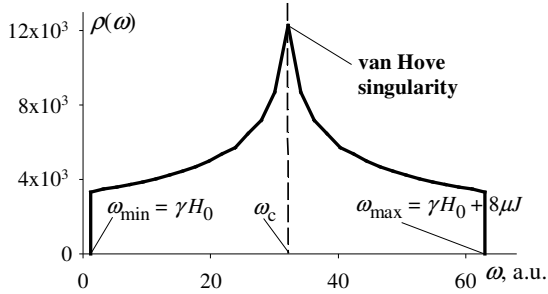

Fig. 2. Density of magnon states for the model discussed in the text. The cusp by $\omega_c = (\omega_{min} + \omega_{max})/2$ is the van Hove singularity common for all 2D models with the cos-like $\omega(\mathbf{k})$ dependence.

The correct excitation spectrum for a real system which we attempt to simulate also contains such a cusp (a real system is discrete at the atomic level) but for frequencies determined by the interatomic distances and thus absolutely unavailable for simulations. This means that the correct spectrum of the model (4) in the frequency region available for micromagnetic simulations is nearly flat as shown by the dashed rectangle in Fig. 1. So in order to obtain correct results for equilibrium system properties using such simulations one should either work in the frequency region where the spectrum is still approximately flat ($\omega \ll \omega_c$) or use a *colored* noise to correct the excitation spectrum of the corresponding lattice model.

## 3. Application examples of Langevin dynamics simulations

### 3.1. Equilibrium magnetic excitations in a nanoelement

The equilibrium magnetic excitation spectrum of nanoelements (one of their most important characteristics) for the given wave vector can be measured using the high-resolution quasielastic Brilloin light scattering (see [10] and Ref. therein). Such experiments provide a highly interesting information about the confinement and quantization of the excitation spectra in such ultrasmall magnetic structures.

To reproduce such spectra using the Langevin dynamics more effort than to obtain the total oscillation power spectrum is required. Namely, after saving all the cell moment trajectories during a simulation time determined by the desired frequency resolution, one has first to perform *spatial* FT to obtain time dependencies $\mathbf{m}(\mathbf{q},t)$ of the spatial Fourier components of the magnetization with the required wave vector $\mathbf{q}$. The subsequent *temporal* FT gives then the power spectrum of these components which can be directly compared with experimental data.

An example of simulation results for a rectangular Py nanodot with lateral sizes 400 x 600 nm and thickness 20 nm discretized into 40 x 60 x 2 cells is presented in Fig. 3.

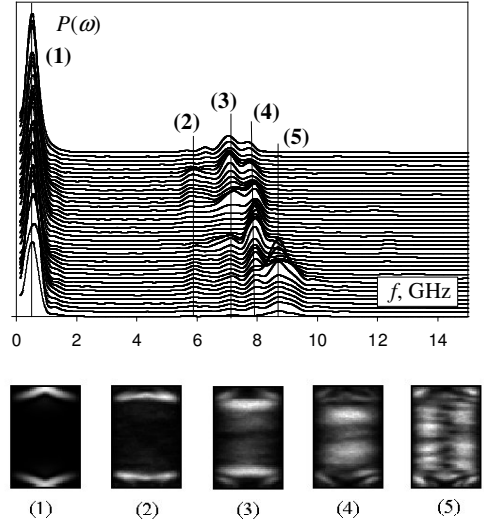

Fig. 3. Oscillation power spectra of a 400 x 500 x 20 nm Py rectangle for wave vectors with $q_x = 0$ and various $q_y$ (upper part) in the in-plane field $H_{ext} = 600$ Oe along the long side (0y-axis) with maps (below) of the spatial distribution of the $m_x$ oscillation power for eigenmodes marked with vertical lines on the spectral graph.

Several peaks with the frequencies which are the same for different wave vectors are observed. This is a natural feature of magnetic excitations in small finite structures where the spatial confinement of an excitation leads to a substantial width of its spectrum in the **q**-space. Detailed comparison with experimental data obtained on larger nanodots will be presented elsewhere [11].

Simulations of this kind also allow to obtain a *spatial* distribution of the oscillation power for each mode by doing temporal FT of the magnetization trajectory of every cell and plotting the 2D spatial dependencies of the spectral power for the corresponding frequency. Such images are shown in Fig. 3 (lower part). It can be seen, e.g., that the lowest mode corresponds to the oscillations of the domain walls between the central part and the 'closure domains' in the 'flower' remanence state.

*3.2. Magnetization dynamics of a square nanodot in a pulsed field: influence of thermal fluctuations*

The most straightforward usage of the Langevin dynamics is the simulation of ultrafast remagnetization processes on a ns-scale in pulsed fields. Such processes could be recently studied by experimentally using Kerr [12] and X-ray [13] microscopy with time and space resolutions enabling direct comparison with micromagnetic simulations.

Here we present an example of such simulations studying the influence of thermal fluctuations on the magnetization dynamics of a Py square element with lateral sizes 1 x 1 µm and thickness 50 nm (discretized in 100 x 100 x 2 cells). External pulse field perpendicular to the layer plane with the trapezoidal time dependence shown in Fig. 3 was applied. The initial magnetization state was assumed to be the closed Landau domain structure.

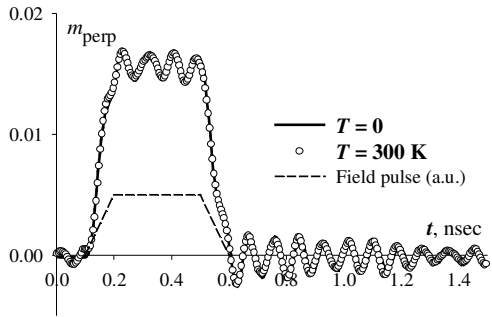

Fig. 4. Comparison of the time-dependencies of the average $m_\perp$-component during and after the field pulse shown as a dashed line.

Time-dependencies of the *averaged* magnetization component perpendicular to the layer plane $m_\perp$ for $T = 0$ and room temperature are shown in Fig. 4. One can see that thermal fluctuations have almost no influence on this dependence (except small fluctuations before the pulse has been applied at $t = 0.1$ ns and long after it was decayed - not shown in Fig. 4).

However, the spatially resolved behaviour of $m_\perp$ is strongly affected by thermal noise. The upper sequence of magnetization images in Fig. 5 represents the time evolution of the magnetization configuration for $T = 0$. It can be seen that domain walls of the Landau structure oscillate with an amplitude and frequency different from that of the domains (which causes a strong contrast along the element diagonals) and that these walls have their own complicate dynamics (last image).

Inclusion of thermal effects leads to the almost complete disappearance of this contrast - only a somewhat brighter homogeneous noisy image can be seen for times corresponding to the pulse plateau in the middle row of Fig. 5, where results for a *single* run are displayed. It was necessary to perform averaging at least over 8 independent runs (noise realizations) to obtain an observable contrast due to domain wall oscillations (image in the middle of the lower series).

Experiments mentioned above [12, 13] are performed using stroboscopic techniques. Hence the simulation results may be used not only to explain experimental data qualitatively (especially for elements with a more complex domain structure), but also to predict the lower boundary for the achievable image contrast for such observations.

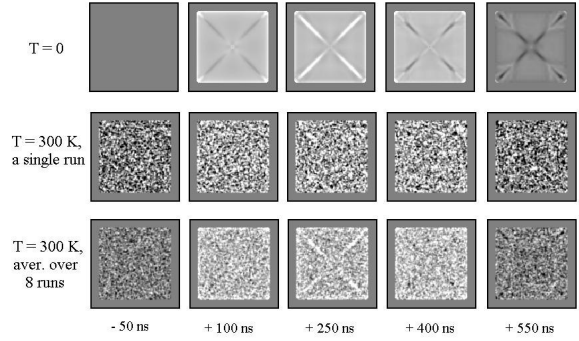

Fig. 5. Spatial maps of the $m_\perp$-component obtained by simulations without thermal noise (upper row), for a single run at $T = 300$ K (middle) and after averaging over 8 runs at $T = 300$ K (last row).

*3.3. Remagnetization dynamics induced by a spin-polarized current injection*

After the theoretical prediction [14] it was soon confirmed experimentally [15] that a spin-polarized current may induce precession and even switching of thin magnetic nanoelements. It was soon realized [16] that simulations of this effect using the Slonczewski torque

$$\Gamma = \frac{a_J}{M_S}[\mathbf{M}\times[\mathbf{M}\times\mathbf{S}]] \qquad (9)$$

(**S** is the spin polarization direction of the current through a layer) within a macrospin model [17] are not sufficient to explain many features of experimental results. Full-scale micromagnetic simulations should address in the first place the following problems: (i) starting from which element size do the domain effects play an important role in the magnetization dynamics and (ii) what is the role of thermal noise in different precession regimes ?

Here we present corresponding preliminary results for a steady-state precession of a 2.5 nm thick square magnetic element with parameters equal to those given in [18]: saturation magnetization $M_S$ = 950 G and the uniaxial anisotropy field along the 0$x$-axis $H_K$ = 500 Oe. Exchange constant (not specified in [18]) was set to $A$ = 1 x $10^{-6}$ erg/cm. All results were obtained for the external field $H_{ext}$ = 1000 Oe along 0$x$ and the spin current strength $a_J$ = 0.4 $M_S$ (**S** also along 0$x$). The lateral mesh size was 2 x 2 nm; results have been proved to be almost independent on the discretization.

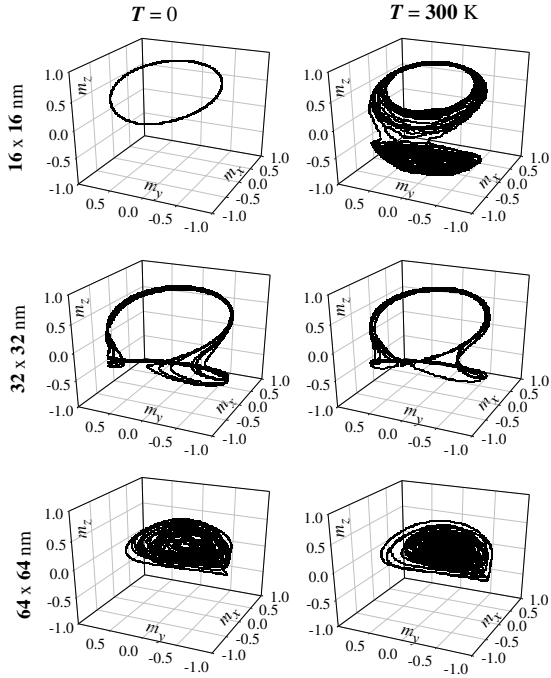

Fig. 6. 3D trajectories of the average element magnetization for the square 16 x 16, 32 x 32 and 64 x 64 nm (see text for details).

We observed already for $T$ = 0 that by increasing the lateral element size $b$ a transition from a single- to multidomain (for $b \approx$ 30 nm) and further to a quasichaotic (at $b \approx$ 60 nm) behaviour occurs (Fig. 6, 7) (detailed description of these transitions is presented in [19]). The striking difference with the results obtained in [18] may be due a much too high exchange constant $A$ used there [20]. We point out that the inclusion of the current magnetic field (not taken into account here and in [18]) should assist the transition to a multi-domain state, due to the inhomogeneity of this field.

The influence of thermal fluctuations depends on the precession regime. For a single-domain state, for which the element size is very small, thermal noise leads to magnetization jumps between the half-spaces above and below the element plane, enabling the 'out-of-plane' precession both with $m_z$ > 0 and $m_z$ < 0 (compare left and right 3D trajectory pictures for a 16 x 16 nm element in Fig. 6). For larger elements with regular multi-domain precession state thermal noise strongly distorts a 'butterfly' trajectory present for $T$ = 0 converting it into a slightly irregular 'bended-8'-type (Fig. 6, picture for 32 x 32 nm). And finally, when the precession is already chaotic, thermal noise has no qualitative influence (Fig. 6, last row) and leads only to further broadening of the magnetization oscillation spectrum (not shown here).

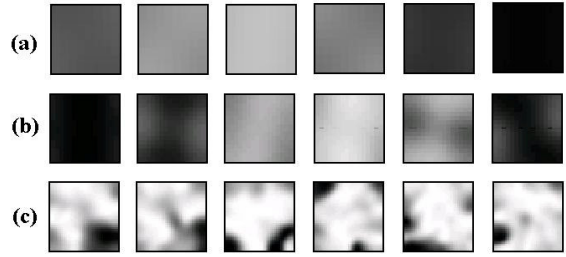

Fig. 7. Typical magnetization patterns (gray-scale maps of $m_x$-projections are shown) during the steady-state precession for elements with the sides $b$ = 16 nm (a), 32 nm (b) and 64 nm (c). Non-homogeneous magnetization pattern for the 32 nm element and chaotic domains for $b$ = 64 nm are clearly visible.

**Acknowledgement.** The authors thank S. Demokritov, J. Miltat and P. Fischer for many useful discussions. This research was partially supported by the Deutsche Forschungsgemeinschaft (project Go 1048/1-P).